\documentclass{article}

\usepackage{amssymb}
\usepackage{amsmath}
\usepackage{amscd}

\def\hs{\hspace{1cm}}

\begin{document}

\begin{titlepage}
\noindent
hep-th/9901164
\hfill ITP--UH--37/98 \\

\vskip 2.0cm

\begin{center}

{\Large\bf NON-LOCAL SYMMETRIES}\\

\medskip

{\Large\bf OF THE CLOSED N=2 STRING}

\vskip 1.5cm

{\Large Klaus J\"unemann\,, \ Olaf Lechtenfeld}

\vskip 0.5cm

{\it Institut f\"ur Theoretische Physik, Universit\"at Hannover}\\
{\it Appelstra\ss{}e 2, 30167 Hannover, Germany}\\
{E-mail: junemann, lechtenf@itp.uni-hannover.de}

\vskip 0.5cm
{\large and}
\vskip 0.5cm

{\Large Alexander D. Popov}

\vskip 0.5cm

{\it Bogoliubov Laboratory of Theoretical Physics}\\
{\it JINR, 141980 Dubna, Moscow Region, Russia}\\
{E-mail: popov@thsun1.jinr.ru}

\end{center}
\vskip 1.5cm

\begin{abstract}
By carefully analysing the picture-dependence of the  BRST cohomology
an infinite set of symmetry charges of the closed $N{=}2$ string is identified.
The transformation laws of the physical vertex operators are shown to coincide
with the linearised non-local symmetries of the Plebanski equation 
(which is the effective field theory of the closed  $N{=}2$ string).
Moreover, the corresponding Ward identities are powerful enough to allow
for a rederivation of the well known vanishing theorem for the tree-level
correlation functions with more than three external legs.
\end{abstract}

\vfill
\end{titlepage}

\section{Introduction}
There are hints that String/$M$-Theory has a large underlying symmetry
whose  improved understanding would certainly be a prerequisite  for
finding a general non-perturbative definition of the theory.
A toy model that might be useful in this context is the closed  $N{=}2$ string
(general references are \cite{OV} - \cite{BL2}).
This is not a theory  of realistic physics since local $(2,2)$ superconformal
symmetry on the world sheet
forces the target space to be a  two complex dimensional Ricci-flat K\"ahler
manifold. However,  it has
the remarkable property that contrary to most other string theories
it possesses only a single massless  scalar degree of
freedom\footnote{Here we
count as degrees of freedom the semi-relative cohomology classes at non-zero
center-of-mass-momentum. For a different viewpoint, see \cite{DL}.}.
Moreover, at tree level all its correlation functions can be calculated,
either explicitly \cite{hipp} or by the more sophisticated method of
Berkovits and Vafa of embedding
the theory into an $N{=}4$ topological string theory \cite{BV}.
Both methods yield the result that all correlators beyond the three-point
function vanish. This is certainly not a coincidence but suggests that
a powerful target space symmetry must be at work.
Such a connection is obvious from the point of view of the effective field
theory, i.e. the field theory
that reproduces {\it all} correlation functions (at tree level) of the string.
In \cite{OV} it has been shown that this is a scalar
theory which describes
the deviation of the  K\"ahler potential from flat space in suitably
chosen coordinates.  The corresponding
equation of motion is known as the Plebanski equation.
It possesses an infinite dimensional symmetry
group (see \cite{PBR} for a description and  references)  which, roughly
speaking, is the loop group of symplectic diffeomorphisms in two real
dimensions.
This symmetry should also be present in the string theory and is surely in
some implicit way contained in the approach of \cite{BV}.
However, Berkovits and Vafa  also stressed the importance of `fleshing out
this symmetry in a more conventional form'. This note is just an attempt
in this direction.

Conventionally, unbroken symmetries in string theory should show up
in the BRST cohomology at ghost number one\footnote{ In this
paper we consider closed strings only and use the
convention that physical states have ghost number two. Other conventions also
exist, but what matters is that symmetries have one unit of ghost
number less than physical states.}.
This is a rather general theorem of string field theory.
The simplest example are target space translations
whose charges can be constructed  from the  cohomology classes
$c \partial X^{\mu}$.
A  more spectacular example is provided by two-dimensional  string theory
in a linear dilaton background
(see \cite{GM} for a review and further references) where an infinite
set of ghost number one cohomology  classes were found at special values
of the momenta. The corresponding charges were shown to form an infinite
dimensional algebra in \cite{W}. Moreover, it is important that  these
results have also  been obtained by matrix-model techniques, which are
completely independent from the BRST approach.

Despite some similarities between the $2D$ string  and the $N{=}2$ string
an analogous
situation will certainly not hold in $N{=}2$ string theory
in an uncompactified target space, for the very
simple reason that the manifest rigid $SU(1,1)$ symmetry in target space
rules out any phenomenon taking place at some distinguished non-zero
value of the center-of-mass momentum. If  symmetries show up in the
BRST cohomology
they can do so only at zero momentum, which is the case we should focus
our attention at.

Theories of closed strings have  the nice property that their Fock space
factorises into  left-moving and   right-moving sectors, which do not talk
to each other and are both isomorphic to the so-called chiral Fock space.
Applying the K\"unneth theorem to the closed string BRST operator
(which is just the sum of left- and right-moving parts) one easily sees that
the BRST cohomology
also factorises. Thus, the most  natural way to construct a ghost number
one cohomology class
is   to combine  a left-moving ghost
number one state (which we know to exist
since this is the ghost number of physical states in the chiral cohomology)
with a right-moving  ghost number zero state, or vice versa.
We therefore arrive at the important conclusion that
{\it  chiral cohomology classes of ghost number zero
signal the existence of symmetry charges for the closed string theory.}
This has been pointed out most clearly in \cite{W}.

Summarising all the above, the first task in analysing the symmetries of the
$N{=}2$ string is to study the chiral cohomology for vanishing momentum and
ghost number. To do this we  have to choose a picture
(see \cite{FMS} and the appendix).
Everything is very simple in the $(-1,-1)$ picture, which in many
aspects is the most natural. It is not hard to show  that in this case
the chiral cohomology at ghost number zero is empty (the same happens in the
$(0,-1)$ and $(-1,0)$ picture)!
This seems
disappointing at first but fortunately it is not the full story: consider
instead the $(0,0)$ picture. Here,  the  ghost number zero cohomology
contains at least the $sl(2)$ invariant ground state of the theory. This state
is certainly BRST invariant but not trivial, for otherwise we had serious
problems with our whole formalism. Thus, one sees  that the zero-momentum
cohomology of the $N{=}2$ string is picture dependent \cite{JL}.
An analogous property for the Ramond sector of the $N{=}1$ string
has been discussed in \cite{BZ}.

One may  wonder what happens if one goes to still higher pictures.
In this case we do not know how to avoid direct computation of the
cohomology which becomes impractical very
quickly\footnote{The usual method to relate the cohomologies
at different pictures by  the picture-changing operation
works for the  $N{=}1$ string at zero momentum
for the absolute, but not for the more important relative cohomology \cite{BZ}.
In the $N{=}2$ theory it  works neither for the relative nor for  the  absolute
cohomology \cite{JL}.}.
The $(0,1)$ and $(1,0)$  pictures can, however,
be treated this way. One finds that their cohomologies contain
two states each! From a technical point of view this is the central result of
our note. Since the BRST cohomology is equipped with a natural multiplication
law \cite{W}, one can take polynomials of the elements in the $(0,1)$
and  $(1,0)$  cohomologies to create more cohomology classes
in higher pictures -- a structure that is reminiscent of Witten's
ground ring in $2D$ string theory. If one is willing to compare picture
number with Liouville momentum,
this analogy becomes a rather close one; recall that both of these
quantum numbers
are the momenta of a scalar field coupled to a background charge.
After a brief general introduction into $N{=}2$ string theory  in section $2$
the detailed description of the cohomology will be presented in section $3$.

Having found zero-momentum states at ghost number zero it is  straightforward
to combine them with ghost number one states
from the same picture to form symmetry charges
of the closed string.
Using the  formalism developed in the context of $2D$ string theory
\cite{WZ,LZ} one then derives
transformation laws for the physical vertex operators which can be compared
with  the symmetries of the Plebanski equation.
This will be done in section $4$.
In section $5$ we derive Ward identities, and it will be shown that
they are strong enough to imply the vanishing of all tree-level
amplitudes with more than three external legs. This constitutes an alternative
proof of the vanishing theorem of \cite{BV}. It also  unambigously shows
that the picture dependence of the
cohomology is not just a bizarre side-effect of BRST quantisation, but
can be used to obtain non-trivial information about the theory.
In  section $6$ the results are summarised, and we make  a couple of
remarks concerning their interpretation.
An  appendix, finally,  contains a  summary of the chiral zero-momentum
BRST cohomologies at ghost number zero and one and a brief
description of the $N{=}2$
ghost system which plays an important role in all what follows.

\section{The $N{=}2$ String}
The  $N{=}2$ string  has a left- and
a right-moving $N{=}2$ superconformal
algebra as constraint algebra. The corresponding ghost system (see appendix)
 has central charge $-6$ implying a critical
dimension $d=4$. The underlying supergravity theory on the world sheet
unfortunately requires the string coordinates to be complex so the target space
is, in fact, two complex dimensional. A free field representation of the 
$N{=}2$ currents is
\begin{equation}
T (z) = - \frac{1}{2} \partial \bar{Z} \cdot \partial Z - \frac{1}{4} \partial
{\psi}^- \cdot \psi^+ - \frac{1}{4} \partial \psi^+ \cdot {\psi}^-,
\nonumber
\end{equation}
\begin{equation}\label{n2currents}
G^{+}(z) = \partial \bar{Z} \cdot \psi^+,\hs G^{-}(z) =
\partial {Z} \cdot {\psi}^-,
\end{equation}
\begin{equation}
J(z) =  \frac{1}{2}
{\psi}^- \cdot \psi^+.\hs \nonumber
\end{equation}
Here $Z^{a}$, $a=0,1$ are the complex string coordinates,
$\bar{Z}^{ \bar a}$  their
complex conjugates and $\psi^{+a}$, ${\psi}^{-\bar{a}}$ the superpartners.
\footnote{Complex conjugation in target space and on the world sheet is
denoted by a bar whereas antiholomorphic operators will be denoted by tilde.
For the fermion fields a $\pm$ index is used instead of a bar. This has the
advantage that the world sheet $U(1)$ charge of any  field equals half the
number of its $\pm$ indices.
The $SU(1,1)$ invariant scalar product is defined through $\eta_{a\bar a}$
with non-vanishing components $ \eta_{1\bar 1} = - \eta_{0\bar 0} = 1$.
For example
$$
\partial {Z} \cdot {\psi}^- = \eta_{a\bar a}
\partial {Z}^a  \cdot {\psi}^{-\bar a} = - \partial {Z}^0  
\cdot {\psi}^{-\bar 0} +\partial {Z}^1  \cdot {\psi}^{-\bar 1}.
$$
Parts of this notation are taken from \cite{BL2}.}

Using the operator product expansions
\begin{equation}\label{ope}
Z^{a} (z) \bar{Z}^{\bar a}(w) \sim -2 \eta^{a\bar a} \ln(z-w) , \hs
\psi^{+a} (z) \psi^{-\bar a}(w) \sim -2 \frac{\eta^{a\bar{a}}}{z-w},
\end{equation}
one may check that the currents satisfy the
$N{=}2$ super Virasoro algebra with central charge $c=6$.
Due to the spectral flow automorphism of this algebra it is no restriction to
consider the NS sector only. Furthermore, one can consider the
additional currents
\begin{equation}\label{n41}
J^{++} = \frac{1}{4} \epsilon_{ab} \psi^{+a} \psi^{+b}, \hs
J^{--} = - \frac{1}{4} \epsilon_{\bar{a}\bar{b}} {\psi}^{-\bar{a}}
{\psi}^{-\bar{b}},
\end{equation}
\begin{equation}\label{n42}
\widehat{G}^+ = \epsilon_{ab} \partial Z^{a} \psi^{+b},\hs
\widehat{G}^- = - \epsilon_{\bar{a}\bar{b}}
\partial \bar{Z}^{\bar a} \psi^{-\bar b}.
\end{equation}
(The antisymmetric $\epsilon$ symbol is defined as 
$\epsilon_{01} = \epsilon_{\bar{0}\bar{1}} = - \epsilon^{01} = 
- \epsilon^{\bar{0}\bar{1}}= 1$.)
Together with the currents (\ref{n2currents}) they satisfy
the small $N=4$ super-conformal algebra \cite{BZ}. In particular the
$J$-currents form an affine  $SU(2)$ algebra:
\begin{align}\label{suz}
J(z) J^{\pm\pm}(w) &\sim \pm \frac{2}{z-w} J^{\pm\pm} (w), \nonumber \\
J^{--}(z) J^{++}(w) &\sim - \frac{1}{(z-w)^2} + \frac{1}{z-w}J(w), \\
J(z) J(w) &\sim \frac{2}{(z-w)^2}. \nonumber
\end{align}

The only physical state of the theory is the massless ground state.
In the $(-1,-1)$ picture the holomorphic part of the corresponding
vertex operator is
\begin{equation}\label{v11}
V_{-1,-1} (k,z) = c e^{-\varphi^+} e^{-\varphi^-} e^{\frac{i}{2}
(k{\cdot}\bar{Z} + \bar{k}{\cdot}{Z})}(z), \hs k\cdot\bar{k}
= 0.
\end{equation}
Using  picture-changed versions of this operator, it is not hard to calculate
the three-point function $A_3 (k_i) = \tilde{A}_3(k_i) \delta(k_1 + k_2 + k_3)$
at tree level:
\begin{equation}
\tilde{A}_3 (k_i) = \big ( k_1{\cdot}\bar{k}_2 - k_2{\cdot}\bar{k}_1  )^2.
\end{equation}
All $N$-point functions with $N>3$ vanish at tree level \cite{hipp,BV}.
These amplitudes are  reproduced by a scalar field $\phi (Z, \bar{Z} )$
with equation of motion
\begin{equation}\label{plf}
\eta^{a\bar a} \partial_a \bar{\partial}_{\bar a} \phi = \frac{1}{2}
\epsilon^{ab} \epsilon^{\bar{a} \bar{b}} \partial_a \bar{\partial}_{\bar a}
\phi \partial_b \bar{\partial}_{\bar b}\phi.
\end{equation}
The geometrical meaning of this equation can be understood
\cite{OV} by considering the Plebanski equation,
\begin{equation}\label{plo}
\epsilon^{ab} \epsilon^{\bar{a} \bar{b}} \partial_a \bar{\partial}_{\bar a}
\Omega \partial_b \bar{\partial}_{\bar b}\Omega = - 2 ,
\end{equation}
which describes the K\"ahler potential $\Omega (Z, \bar{Z} )$ of a
Ricci-flat K\"ahler metric in
a suitably chosen coordinate system. The field $\phi$ parametrises  deviations
of $\Omega$ from flat space, since inserting the expression
\begin{equation}
\Omega = \eta_{a\bar a} Z^a \bar{Z}^{\bar a} + \phi(Z, \bar{Z} )
\end{equation}
into (\ref{plo}) yields (\ref{plf}).
An obvious symmetry of these equations  are the  usual K\"ahler
transformations.

Interestingly the Plebanski equation is equivalent
to the consistency condition $[ {\cal L}_0, {\cal L}_1] = 0$ of the  linear
system \cite{PBR}
\begin{equation}
{\cal L}_0 =  \partial_0 + \lambda W_0, \hs {\cal L}_1 =  \partial_1 +
\lambda W_1
\end{equation}
with
\begin{equation}\label{va}
W_a =\epsilon^{\bar{a} \bar{b}} \partial_a \bar{\partial}_{\bar a}
\Omega \bar{\partial}_{\bar b}
\end{equation}
and an arbitrary complex  parameter $\lambda$.
This structure is familiar from the theory of integrable models
and usually leads to a large  symmetry which will be further
explored in subsection  $4.3$. It is the  main purpose  of this paper to
investigate how these symmetries are realized in the $N{=}2$ string.

\section{Chiral BRST cohomology}
As has been mentioned in the introduction and will be further explained in
section  $4$, an important technical tool to study unbroken
symmetries in string
theory is the chiral BRST cohomology (i.e. the cohomology of only the
left-moving part of the Fock space).
There exists a powerful method to solve the cohomology problem
for non-zero momentum \cite{FGZ,LZ2}. The result of this analysis for
the $N{=2}$ string \cite{JL} is that the chiral, relative (see below)
cohomology contains precisely one physical state in each picture.  This state
has ghost number one, and in the $(-1,-1)$ picture it is represented by the
vertex operator (\ref{v11}).
Unfortuntely, the method of \cite{FGZ,LZ2} fails for vanishing momentum.
This is unimportant if one is only interested in the spectrum of the theory,
since the behaviour of the dynamical degrees of freedom at isolated points
in momentum space is irrelevant. For the symmetry structure, however,
the zero-momentum cohomology is very important.

We do not know of a  systematic method to completely determine  the
cohomology for zero momentum except  for explicit computation, which is the
subject of the present chapter.
For this the following elementary and  well-known observation
is  crucial: The spectrum of the zero modes $L_0$ and $J_0$
of the bosonic currents is discrete so  that representatives of non-trivial
cohomology classes can always be chosen to be annihilated by $L_0$ and $J_0$.
To prove this one uses the relations
\begin{equation}
\{ Q, b_0 \} = L_0 , \hs \{ Q , b'_0\} = J_0
\end{equation}
to show that a BRST-invariant state with non-vanishing eigenvalue is always
trivial. These relations imply that the space of states involved
in the cohomology problem can be further constrained by the requirement that
all states be annihilated by $b_0$ and $b'_0$. We therefore need to
consider only the relative Fock space $F_{rel}$, defined as
\begin{equation}\label{rfs}
F_{rel} := \Big \{ |\psi \rangle \;\; \Big | \;\; L_0 |\psi \rangle 
= J_0 |\psi \rangle
= b_0 |\psi \rangle = b'_0 |\psi \rangle = 0 \Big  \}.
\end{equation}
The BRST cohomology of this space is called relative cohomology; in this
section, however, the term `relative' will be dropped.
What happens to the cohomology when the $b_0$ and $b'_0$ conditions are
relaxed is described in great detail in section $3$ of \cite{WZ}.

Moreover, we only need to consider pictures $\pi^{\pm}\geq -1$. The other
half of the cohomology  can  be found by
Poincar\'e duality (see section 3 of \cite{LZ2}, for example) once
the first half is known.

\subsection{Ghost number zero}
Let us first consider the $(-1,-1)$ picture.
At ghost number\footnote{As in
\cite{B} ghost number is defined to commute with picture number,
see also the appendix.} zero it is simply impossible to write down a
state obeying
the conditions in (\ref{rfs}). The relative Fock space  is empty and
so is the cohomology.
Similarly, one shows that there is no cohomology at ghost number zero
in the $(0,{-}1)$ and $({-}1,0)$ picture.

In the $(0,0)$ picture the situation is different. The relative Fock space
contains two candidate states at zero ghost number, namely
\begin{equation}
|\; 0,0 , k=0 \;\rangle \hs {\rm and}\hs  c_1 b'_{-1}|\; 0,0 , k=0 \;\rangle.
\end{equation}
The first of these states is just the $sl(2)$ invariant ground state which
is BRST invariant, whereas the second state  is not invariant.
Since the relative Fock space at ghost number $-1$ is empty in this picture,
we need not worry about the image of $Q$ and have thus proven that the ground
state (or the unit operator in the language  of vertex operators) spans the
ghost number zero  cohomology.

It is instructive also to consider the $(-1,1)$ picture. Candidate states are
\begin{equation}
d^{-\bar a}_{-1/2} d^{-\bar b}_{-1/2} |\; {-}1,1 , k=0 \;\rangle, \hs
c_1 b'_{-1} d^{-\bar a}_{-1/2} d^{-\bar b}_{-1/2} |\; {-}1,1 , k=0 \;\rangle ,
\end{equation}
where the Fourier modes of the matter fermions $\psi^{\pm}$ appear:
\begin{equation}
i \psi^{+a} = \sum_{r \in \mathbb{Z} + 1/2} d^{+a}_r z^{-r-1/2}, 
\hs i \psi^{-\bar{a}} = \sum_{r \in \mathbb{Z} + 1/2}
d^{-\bar{a}}_r z^{-r-1/2}.
\end{equation}
A small calculation shows that the combination
\begin{equation}\label{sz}
(1-c_1 b'_{-1}) d^{-\bar a}_{-1/2} d^{-\bar b}_{-1/2}|\; {-}1,1 , k=0\;\rangle
\end{equation}
is invariant. Again there is no state at ghost number $-1$ so that (\ref{sz})
spans the cohomology.
The corresponding vertex operator is
\begin{equation}\label{a}
A(z) := (1 - c b') J^{--} e^{\varphi^+} e^{-\varphi^-} (z).
\end{equation}
In exactly the same way one shows that the cohomology in the $(1,-1)$ picture
is represented  by
\begin{equation}\label{ta}
\hat{A}(z) := (1 + c b') J^{++} e^{-\varphi^+} e^{\varphi^-} (z).
\end{equation}
The operators $A$ and $\hat{A}$ are  nothing but spectral flow
operators with spectral parameter $1$ and $-1$. They induce an isomorphism
between the cohomologies at picture $(\pi^+,\pi^-)$ and $(\pi^+-1,\pi^-+1)$.
To show this one uses the fact  that the BRST cohomology possesses
a natural multiplication
rule \cite{W}: given two BRST invariant but  non-trivial vertex
operators $O_1$ and $O_2$, their normal ordered product
\begin{equation}\label{mult}
\big (O_1 \cdot O_2\big ) (w) 
= \oint_w \frac{dz}{2 \pi i} \frac{1}{z-w} O_1(z) O_2(w)
\end{equation}
defines a new cohomology class. This product\footnote{
There is a little problem with this product if one considers general
vertex operators with momentum $k$, since then the OPE between the vertex 
operators typically contains singularities of the form $(z-w)^{k_1 \cdot k_2}$.
Therefore the product makes sense only between operators whose momenta are
constrained to have integer scalar product. In this paper we only consider the
case where at least one operator involved has zero momentum and so  this
difficulty disappears.}
has the extremely important
property that on cohomology classes it is graded commutative and
associative \cite{LZ}.
Using (\ref{suz}) one finds that the product of $A$ and $\hat A$ is
\begin{equation}
A \cdot \hat{A} = 1 \hs \Rightarrow \hs \hat{A} = A^{-1}.
\end{equation}
Multiplication by $A$ is thus an invertible map between the cohomologies
at $(\pi^+,\pi^-)$ and $(\pi^+-1,\pi^-+1)$ which shows that they are
isomorphic. This is how the spectral flow automorphism of the $N{=}2$
super Virasoro algebra acts on its BRST cohomology.

Let us now turn to the $(0,1)$ and $(1,0)$ pictures. What do we expect to find?
Recall that there exist  picture-changing
operators $X^{\pm}$. They  act on physical states by the above
multiplication rule (\ref{mult}). According to
Friedan, Martinec and Shenker \cite{FMS} they are constructed rather
ingeniously as
$X^{\pm}=\{Q , \xi^{\pm} \}$ which are non-trivial since the zero modes of
$\xi^{\pm}$ are not part of  the theory (bosonization of the $\beta$  ghosts 
involves only $\partial \xi^{\pm}$).
Their explicit form is
\begin{equation}
X^{\pm} = - c \partial \xi^{\pm} + \big ( G^{\pm} - 4
\gamma^{\pm} b \pm 4 \partial \gamma^{\pm} b' \pm 2 \gamma^{\pm} \partial
b'\big )  e^{\varphi^{\mp}}.
\end{equation}
$X^+$ has picture number $(1,0)$ and $X^-$ has $(0,1)$.
The states $X^{\pm} (z=0) |0\rangle$ are  ordinary
cohomology classes with ghost number zero and vanishing momentum.
There exists, however, an alternative way to construct a cohomology class
in the $(0,1)$ picture, say:  just consider the operator 
$Y^- := A \cdot X^+$ \cite{BL1}. Its explicit form is
\begin{align}
Y^- &= \big [ c\beta^+ e^{\varphi^{+}} - 4 ( 1 - c b' )
(b + \frac{1}{2} \partial b') \gamma^+ e^{\varphi^{+}} + 4 b'
\gamma^+ \partial  e^{\varphi^{+}} \big ] J^{--}\nonumber \\
& + ( 1 - c b' )e^{\varphi^{+}} \widehat{G}^-.
\end{align}
The important observation is  that {\it $Y^-$ is  BRST inequivalent to
$X^-$}. If they were equivalent some linear combination of $X^-$ and $Y^-$
had to be trivial. But this cannot be the case since one may easily convince
oneself that the relative Fock space in the $(0,1)$ picture at ghost number
$-1$ consists only of the state $b'_{-1} d^{-\bar{a}}_{-1/2}
|\; 0,1 , k=0 \;\rangle$. Obviously the image of this state under $Q$ is not
equal to a linear combination of $X^-(0)|0\rangle$ and $Y^-(0)|0\rangle$,
which shows that they are
inequivalent. Alternatively, one could find  $X^-$ and $Y^-$ by
explicitly writing down a basis of the relative Fock space at ghost number zero
in this picture and look for  BRST-invariant
combinations. In this way one can prove that there are no further cohomology
classes in this picture besides $X^-$ and $Y^-$.
In complete analogy one shows that the cohomology in the $(1,0)$ picture
is represented by $X^+$ and $Y^+ := X^- \cdot A^{-1}$.

It is instructive to work out how these operators act on physical states.
Let us start with the vertex operator in the $(-1,-1)$ picture,  
given in (\ref{v11}), and define vertex operators in higher pictures as
\begin{equation}\label{vo}
V_{\pi^+,\pi^-} (k) = (X^+)^{\pi^+ + 1} \cdot (X^-)^{\pi^- + 1}
\cdot V_{-1,-1} (k).
\end{equation}
To see what $A$ and $A^{-1}$ do, consider the operators
\begin{align}
V_{-1,0} (k) = c k{\cdot}\psi^-   e^{-\varphi^-} e^{\frac{i}{2}
(k{\cdot}\bar{Z} + \bar{k}{\cdot}{Z})}, \nonumber \\
V_{0,-1} (k) = c \bar{k}{\cdot}\psi^+   e^{-\varphi^+} e^{\frac{i}{2}
(k{\cdot}\bar{Z} + \bar{k}{\cdot}{Z})}.
\end{align}
Using the explicit expressions (\ref{a}) and (\ref{ta}) it is easy to
check that
\begin{equation}\label{aaction}
A\cdot V_{0,-1}(k) = h(k) V_{-1,0}(k), \hs A^{-1}\cdot V_{-1,0}(k) 
= h(k)^{-1} V_{0,-1} (k)
\end{equation}
with $h(k)$ defined as
\begin{equation}
h(k) = \frac{\bar{k}^0}{k^1} = \frac{\bar{k}^1}{k^0}.
\end{equation}
Note that $|h| = 1$ so that $h^{\ast} = 1/h$.
This particular function of the momenta features prominently in \cite{BV}
(see also \cite{Parkes})
and will also be very important in what follows. In position space $h$
translates into a non-local expression, indicating that we are on the right
track to discover the non-local symmetries of the Plebanski equation
on the string theory side.
Using the commutativity of the product (\ref{mult}) and the definition 
(\ref{vo}) one sees that $A$ and $A^{-1}$ act on all higher vertex operators 
as in equation (\ref{aaction}).

Ghost number zero cohomology classes in higher pictures can now be
constructed by simply
considering positive powers of $X^{\pm}$ and integer  powers of $A$.
For  a given picture  $(\pi^+, \pi^-)$ one can write
down $\pi^+ + \pi^- + 1$ operators:
\begin{equation}\label{coh}
{\cal O}_{\pi^+,\pi^-, n} : =  (X^+)^{\pi^++n} \cdot (X^-)^{\pi^--n}\cdot A^n
, \hs n= - \pi^+ , ..., \pi^-.
\end{equation}
The range of $n\in \mathbb{Z}$ is restricted  because negative powers of
$X^{\pm}$ do not exist.
The operators ${\cal O}_{\pi^+,\pi^-, n}$ are all non-trivial. To see this
assume that some linear combination  of the
${\cal O}_{\pi^+,\pi^-, n}$ were BRST trivial, i.e.
\begin{equation}
\sum_n \alpha_n {\cal O}_{\pi^+,\pi^-,n} = \{Q , \Lambda\}
\end{equation}
for some $\Lambda$ and suitably chosen coefficients $\alpha_n$.
Then its product with any vertex operator $V(k)$ necessarily had to
be trivial, as well. On the other hand one has, using (\ref{aaction}),
\begin{equation}
\Big (\sum_n \alpha_n {\cal O}_{\pi^+,\pi^-,n}\Big ) \cdot V_{0,0}(k) 
= \Big ( \sum_n \alpha_n  {h(k)}^n \Big ) V_{\pi^+,\pi^-}(k).
\end{equation}
The right hand side  can only be BRST trivial if the sum $\sum_n \alpha_n
h(k)^n$ vanishes
for any  momentum  $k$ with $k{\cdot}\bar{k}=0$. This can certainly not
be the case,
which proves that any of the operators in (\ref{coh}) represents a distinct
cohomology class. This does not prove, however, that
the operators (\ref{coh})
span the full cohomology. It may  well be that there exist additional
cohomology classes that cannot be constructed this way.

In order to  summarize the above findings let us introduce the notation
$\pi = \pi^+ + \pi^- $.
Then the following facts
about the ghost number  zero
cohomologies at vanishing momentum hold:
\begin{itemize}
\item
The cohomologies at picture $(\pi^+,\pi^-)$ and
$(\widehat{\pi}^+,\widehat{\pi}^-)$ are isomorphic
if $\pi = \widehat{\pi}$
(this is just the spectral flow automorphism and is  also  true for
other ghost numbers and any momentum).

\item
There is no cohomology for $\pi =  -2$ or $\pi = -1$.

\item
There is one cohomology class for $\pi =0$. 
For $(\pi^+,\pi^-) = (-n,n)$ this class is represented by the operator $A^n$.

\item
There are two classes for $\pi =1$. For $(\pi^+,\pi^-) = (1-n,n)$
they are represented by $X^+ \cdot A^{n}$ and $X^- \cdot A^{n-1}$.

\item
For $\pi > 1 $ the cohomology contains the states of (\ref{coh}).
All these states  represent  distinct cohomology classes but may not
exhaust the full cohomology.

\end{itemize}
Analogous statements for $\pi<-1$ can be obtained by Poincar\'e duality.

\subsection{Ghost number one}
The zero momentum cohomology at ghost number one contains one state in the
$(-1,-1)$ picture and two states in the  $(-1,0)$ and $(0,-1)$  pictures 
respectively \cite{JL}.
In this paper, however, we are mainly interested in  the case  $\pi \geq 0$.
In the $(0,0)$  picture, the cohomology at
ghost number one is spanned by the four operators \cite{Bi}
\begin{equation}\label{mom}
-i{\cal P}^a_{0,0} = c \partial Z^{a} - 2 \gamma^- \psi^{+a}, 
\hs -i \bar{\cal{P}}^{\bar a}_{0,0} =
c \partial \bar{Z}^{\bar a} - 2 \gamma^+ \psi^{-\bar a}.
\end{equation}

Multiplication with the  operators of equation
(\ref{coh}) yields new cohomology classes in higher pictures,
\begin{align}\label{coh1}
{\cal P}^a_{\pi^+,\pi^-, n} := {\cal O}_{\pi^+,\pi^-, n}
\cdot {\cal P}^a_{0,0}, \nonumber \\
\bar{{\cal P}}^{\bar a}_{\pi^+,\pi^-, n} := {\cal O}_{\pi^+,\pi^-, n} \cdot
\bar{{\cal P}}^{\bar a}_{0,0}.
\end{align}
For a given picture $(\pi^+,\pi^-)$ these are $4( \pi + 1)$ states. To
show that
they all represent different cohomology classes we need the so-called
bracket operation \cite{LZ}, defined as
\begin{equation}\label{bracket}
\{ O_1, O_2 \} (w) := \oint_w \frac{dz}{2 \pi i} O_1^{(1)} (z) O_2(w)
\end{equation}
with
\begin{equation}\label{1}
O^{(1)} (w) := \oint_w \frac{dz}{2 \pi i} b(z) O(w).
\end{equation}
$O_1$ and $O_2$ are arbitrary operators and $b$ is the anti-ghost field.
In \cite{LZ} it has been shown that on cohomology classes the bracket operation
is graded commutative although the operators $O_1$ and $O_2$ enter quite
differently into the definition (\ref{bracket}). Moreover, it acts as a 
graded derivation on  the normal ordered product in the cohomology:
\begin{equation}\label{rule}
\{ O_1, O_2\cdot O_3 \} = \{ O_1, O_2\} \cdot O_3 + (-)^{(|O_1| -1)|O_2|}
O_2 \cdot \{O_1, O_3\}.
\end{equation}
As usual, the symbol $|O|$ is zero if $O$ is a bosonic operator and one
if it is fermionic. The BRST cohomology together with the normal ordered
product and the bracket operation has been called `Gerstenhaber algebra'
in \cite{LZ} where more details and references to the original work
of Gerstenhaber can be found.
Obviously the bracket carries ghost number $-1$, so it is the appropriate
operation for the ghost number one states (\ref{coh1}) to define an
operator map within each sector of fixed ghost number.
The operators in (\ref{mom}) act on the physical vertex operators as momentum
operators,
\begin{align}
\{ {\cal P}^a_{0,0} , V_{\pi^+,\pi^-} (k) \} &= k^a V_{\pi^+,\pi^-} (k),
\nonumber \\
\{ \bar{\cal{P}}_{0,0}^{\bar a}, V_{\pi^+,\pi^-} (k) \} &= 
\bar{k}^{\bar a} V_{\pi^+,\pi^-} (k).
\end{align}
This can easily be shown by performing the computation explicitly in the
$(-1,-1)$ picture and then using equations (\ref{vo}), (\ref{rule}) and the
relations
\begin{equation}
\{ {\cal P}^a_{0,0} , X^{\pm} \} = 
\{ {\cal P}^a_{0,0}, A \} = \{ \bar{{\cal P}}^{\bar a}_{0,0}, X^{\pm}\} =
\{ \bar{{\cal P}}^{\bar a}_{0,0}, A\} = 0,
\end{equation}
which obviously hold
since $X^{\pm}$ and $A$ are operators with vanishing momentum.
Similarly, one  shows that
\begin{align}
\{{\cal P}^a_{\pi^+,\pi^-,n} , V_{\widehat{\pi}^+,\widehat{\pi}^-} (k) \} =
h(k)^n k^a V_{\pi^+ + \widehat{\pi}^+,\pi^- + \widehat{\pi}^-}(k),
\nonumber \\
\{ \bar{{\cal P}}^{\bar a}_{\pi^+,\pi^-,n}, V_{\widehat{\pi}^+,\widehat{\pi}^-}
(k) \} = h(k)^n  \bar{k}^{\bar a} V_{\pi^+ + \widehat{\pi}^+,\pi^- + 
\widehat{\pi}^-} (k).
\end{align}
With the same argument that established the BRST inequivalence
of the operators in equation (\ref{coh}) one now proves that
the operators in (\ref{coh1}) all represent distinct cohomology classes.
But, as in the ghost number zero case, there may well be more cohomology
in higher pictures.

\subsection{Ghost number two and higher}
There are plenty of cohomology classes at arbitrarily high ghost number
when $\pi^+ + \pi^-
\geq -1$. They can be obtained by acting with $X^{\pm}$ and $A$ on the states
$A_N$, $B_N$ and $C_N$ of equation (15) in \cite{JL}, but their role
within the symmetry structure of the theory is not clear to us.

\section{Symmetries}
We can now use the results of the previous section to study the
symmetries of the $N{=}2$ string. The machinery to systematically analyse
symmetries and Ward identities of a theory -- once its BRST cohomology
is known -- has been developed in the context of $2D$ string theory.
For completeness we briefly review some of the material following \cite{WZ}.

\subsection{Some generalities}
A current with components $J_z$ and $J_{\bar z}$ in a two dimensional theory
(complex coordinates $z$ and $\bar z$) is conserved, i.e.
satisfies $\bar{\partial} J_z + \partial J_{\bar z} = 0$, when the one-form
\begin{equation}
\Omega^{(1)} =  J_z dz - J_{\bar z} d\bar{z}
\end{equation}
is closed. The corresponding charge
\begin{equation}
{\cal A} = \oint_{\cal C} \Omega^{(1)}
\end{equation}
is conserved if it has the same value for contours $\cal C$ and
${\cal C}'$ that are homologous, i.e. are the boundaries of some surface $M$,
$\partial M = {\cal C} - {\cal C}' $. Current conservation implies charge
conservation by Stokes' Theorem.

In BRST quantisation these relations are required to  hold  only up
to BRST commutators. Current conservation then reads\footnote{
Depending on whether two operators  are both
fermionic or not,  one must consider their
anti-commutator or commutator. In this section we generally denote this by
$[\cdot , \cdot ]_{\pm}$. It should always be clear from the context which
is meant. We use this notation in order to avoid confusion with
the Gerstenhaber bracket which, strictly speaking, is not an anticommutator.}
\begin{equation}
d \Omega^{(1)} = \big [ Q , \Omega^{(2)} \big{]}_{\pm}
\end{equation}
for some two form $\Omega^{(2)}= \Omega^{(2)}_{z\bar z} dz \wedge d\bar{z}$.
BRST invariance of the charge requires
$[ Q, \Omega^{(1)} ]_{\pm} = d \Omega^{(0)}$.
Applying $Q$ to this relation implies that $ [ Q, \Omega^{(0)} ]_{\pm}$
is a constant. Furthermore, this constant must vanish, since otherwise
the unit operator were BRST trivial. Summarising, we
have the descent equations
\begin{align}\label{de}
\big [ Q, \Omega^{(0)} \big{]}_{\pm} &=0, \nonumber \\
\big [ Q, \Omega^{(1)} \big{]}_{\pm} &= d \Omega^{(0)},\\
\big [ Q, \Omega^{(2)} \big{]}_{\pm} &= d \Omega^{(1)}.\nonumber
\end{align}
Moreover, the whole formalism is unaffected by the replacements
\begin{align}
\Omega^{(0)} &\rightarrow \Omega^{(0)} +  \big [ Q, \alpha \big{]}_{\pm} ,
\nonumber \\
\Omega^{(1)} &\rightarrow \Omega^{(1)} + d\alpha ,
\end{align}
where $\alpha$ is some form of degree zero.
The equations for $\Omega^{(0)}$ are precisely the defining relations for the
BRST cohomology. Taking as $\Omega^{(0)}$ some cohomology class of ghost
number $g$ one can then construct the forms $\Omega^{(1)}$ and $\Omega^{(2)}$
of ghost numbers $g-1$ and $g-2$. It is  most natural   to choose
$\Omega^{(0)}$ to have ghost number one. This results in a charge of ghost
number  zero that can map physical states to physical states.
In principle one could also
consider charges of different ghost number (see \cite{V} for a discussion), 
but they would annihilate the physical states.

The fact that symmetries sit at ghost number one can also be seen in
a string field
approach, in which the equation of BRST invariance $Q | \Psi \rangle = 0$
can be regarded as the linearized equation of motion for the string field
$| \Psi \rangle$. Schematically, gauge symmetries have the form
\begin{equation}
\delta | \Psi \rangle = Q | \Lambda \rangle + | \Psi \ast \Lambda \rangle +
{\rm higher\;order\;terms}.
\end{equation}
Since the string field has ghost number two (it contains the physical states)
the symmetry parameter $| \Lambda \rangle$ has ghost number one.
If $| \Lambda \rangle$ is not killed by $Q$ we have a gauge symmetry that
starts with a field-independent term. Such a symmetry can often be used to
simply gauge away some fields  (see section 4 of \cite{P}). It is  not
an unbroken symmetry of the chosen background.  Moreover,
a BRST-trivial  parameter $| \Lambda \rangle = Q | \Sigma \rangle$
leads to a trivial symmetry  that cannot be used to derive conserved currents.
We thus see again
that ghost number one cohomology is the right tool for studying symmetries
in string theory\footnote{We have neglected the effects of the  anti-ghost
zero modes here. Taking them into account properly leads the
semi-relative cohomology. We leave it to future work to do this for
the $N{=}2$ string.}.

As a simple example let us consider target space translations in bosonic
string theory. The only ghost number zero chiral cohomology class is the
unit operator, which we may take as the right-moving piece of the closed
string
cohomology. As left-moving piece we must take a ghost number one state that
also has vanishing momentum -- the only candidate is $c \partial X^{\mu}$.
The forms $\Omega^{(i)}$  take the form (suppressing the right-moving unit
operator)
\begin{equation}\label{bosstr}
\Omega^{(0)} = c \partial X^{\mu}, \hs \Omega^{(1)} = \partial X^{\mu} dz, \hs
\Omega^{(2)}=0.
\end{equation}
The charge is just the center-of-mass momentum operator 
$P^{\mu} = \oint \frac{dz}{2\pi}\partial X^{\mu}$.

\subsection{Transformation laws of physical states}
In this section the formalism just described will be applied
to the $N{=}2$ string. To construct the symmetry charges we first have
to specify the ghost number one cohomology classes $\Omega^{(0)}$ of the
closed strings. Without loss of generality one can take a chiral
cohomology class of ghost number one as left-mover and one of ghost number zero
as right-mover (denoted by $\,\tilde{}\;$). Exchanging left- and right-movers
does not lead to anything new. Moreover left- and right-movers are chosen 
to have the same picture number. Using the expressions in equations (\ref{coh}) 
and (\ref{coh1}) leads to
\begin{equation}\label{coh2}
\Omega^{a(0)}_{\pi^+,\pi^-,m,n} = {\cal P}^a_{\pi^+,\pi^-,m}(z)
\tilde{\cal O}_{\pi^+,\pi^-,n} (\bar z), \hs m,n = - \pi^+, ... ,\pi^-.
\end{equation}
An analogous zero form can be constructed from $\bar{\cal P}^{\bar a}$.
$\Omega^{(1)}$ and  $\Omega^{(2)}$ can be found
using the fact that, given a chiral cohomology class $V$, the `current'
$V^{(1)}$ defined as in equation (\ref{1}) BRST-transforms into the derivative
of $V$,
\begin{equation}
\big [ Q , V^{(1)}\big{]}_{\pm} = \partial V.
\end{equation}
It is then not hard to check that $\Omega^{(0)}$ together with\footnote
{ For notational
simplicity  the vector indices $a$ and $\bar a$ are suppressed in eqs.
(\ref{o1}) -- (\ref{comav}).}
\begin{equation}\label{o1}
\Omega^{(1)}_{\pi^+,\pi^-,m,n} = - {\cal P}^{(1)}_{\pi^+,\pi^-,m}(z)
\tilde{\cal O}_{\pi^+,\pi^-,n} (\bar z)
dz +  {\cal P}_{\pi^+,\pi^-,m}(z) \tilde{\cal O}^{(1)}_{\pi^+,\pi^-,n}
(\bar z) d\bar{z}
\end{equation}
and
\begin{equation}
\Omega^{(2)}_{\pi^+,\pi^-,m,n} =  {\cal P}^{(1)}_{\pi^+,\pi^-,m}(z)
\tilde{\cal O}^{(1)}_{\pi^+,\pi^-,n}(\bar z) dz \wedge d\bar{z}
\end{equation}
satisfy the descent equation.

Let us now compute how the symmetry charges
\begin{align}
{\cal A}_{\pi^+,\pi^-,m,n} &= \oint \frac{dz}{2\pi i}
{\cal P}^{(1)}_{\pi^+,\pi^-,m}(z) \tilde{\cal O}_{\pi^+,\pi^-,n} (\bar z)
\nonumber \\
&- \oint \frac{d\bar z}{2\pi i}  {\cal P}_{\pi^+,\pi^-,m}(z)
\tilde{\cal O}^{(1)}_{\pi^+,\pi^-,n} (\bar z)
\end{align}
act on the closed string vertex operators
\begin{equation}
V_{\widehat{\pi}^+,\widehat{\pi}^-}(k,z,\bar{z}) = V_{\widehat{\pi}^+,
\widehat{\pi}^-}(k,z)
\tilde{V}_{\widehat{\pi}^+,\widehat{\pi}^-}(k,\bar{z})
\end{equation}
(the holomorphic and antiholomorphic pieces are defined in eq. (\ref{vo})).
The calculation can be done by rewriting the contour integrals in terms of
the product (\ref{mult}) and the Gerstenhaber bracket (\ref{bracket})
\footnote{
A completely analogous computation which is described in some more detail 
is the derivation of equation $(5.19)$ in \cite{WZ}.}.
\begin{align}\label{comav}
&\delta_{\pi^+,\pi^-,m,n}V_{\hat\pi^+,\hat\pi^-}(k,w,\bar w):=
[{\cal A}_{\pi^+,\pi^-,m,n},  V_{\widehat{\pi}^+,\widehat{\pi}^-}(k,w,\bar{w}) ]
\nonumber \\  &=
\oint_w \frac{dz}{2\pi i}  {\cal P}^{(1)}_{\pi^+,\pi^-,m}(z)
V_{\widehat{\pi}^+,\widehat{\pi}^-}(w) \tilde{\cal O}_{\pi^+,\pi^-,n} (\bar z) 
\tilde{V}_{\widehat{\pi}^+,\widehat{\pi}^-}(\bar{w})\nonumber \\
&+ \oint_{\bar w} \frac{d\bar z}{2\pi i}  {\cal P}_{\pi^+,\pi^-,m}(z)
V_{\widehat{\pi}^+,\widehat{\pi}^-}(w)
\tilde{\cal O}^{(1)}_{\pi^+,\pi^-,n} (\bar z)
\tilde{V}_{\widehat{\pi}^+,\widehat{\pi}^-}(\bar{w})
\nonumber \\
&= \{ {\cal P}_{\pi^+,\pi^-,m}, V_{\widehat{\pi}^+,\widehat{\pi}^-} \}(w) \;
\tilde{\cal O}_{\pi^+,\pi^-,n} {\cdot} \tilde{V}_{\widehat{\pi}^+,
\widehat{\pi}^-}(\bar{w}) \nonumber \\
&+ {\cal P}_{\pi^+,\pi^-,m} {\cdot}
V_{\widehat{\pi}^+,\widehat{\pi}^-} (w)\;
\{ \tilde{\cal O}_{\pi^+,\pi^-,n} ,
\tilde{V}_{\widehat{\pi}^+,\widehat{\pi}^-} \} (\bar{w}) .
\end{align}
The  term in the fourth line is known from section $3$ and the last term
vanishes because of ghost number counting. To see this consider the expression
${{\cal P}_{\pi^+,\pi^-,m}\cdot V_{\widehat{\pi}^+,\widehat{\pi}^-}(k) (w)}$ 
which is an operator
with momentum $k\neq 0$ and ghost number two. However, from the general
analysis \cite{JL} it is known that there is no chiral cohomology at this
ghost number. We thus proved that this operator
must be BRST trivial. The final result then reads
\begin{align}\label{trafo}
&\delta_{\pi^+,\pi^-,m,n}^aV_{\hat\pi^+,\hat\pi^-}(k,w,\bar w)=
[{\cal A}^a_{\pi^+,\pi^-,m,n}, 
V_{\widehat{\pi}^+,\widehat{\pi}^-}(k,w,\bar{w}) ] 
\nonumber \\
&= h(k)^{m+n}
k^a  V_{\pi^{+}+\widehat{\pi}^{+}, \pi^{-}+\widehat{\pi}^{-} }(k,w,\bar{w}).
\end{align}
An analogous expression with $k^a$ replaced by $\bar{k}^{\bar a}$ also holds.

\subsection{Symmetries of the Plebanski equation}
In order to propose a possible interpretation of  the transformations
(\ref{trafo}),
we study in this subsection the symmetries of the Plebanski
equations (\ref{plf}) and  (\ref{plo}), hoping  to find
something  resembling (\ref{trafo}). The symmetry structure of self-dual
gravity has been
intensively investigated (see \cite{PBR} for further references)
and can be described in a precise and mathematically
rigorous way in terms of twistor theory. However, this formalism is
unnecessarily abstract for our purposes, and  we prefer  a more
pedestrian but explicit  approach. So instead of trying to describe the
symmetries in full generality we only work out specific  transformation
laws, keeping in mind that they are part of  a deep underlying mathematical
structure.

Recall that the Plebanski equation
\begin{equation}\label{pl2}
\epsilon^{ab} \epsilon^{\bar{a} \bar{b}} \partial_a \bar{\partial}_{\bar a}
\Omega\, \partial_b \bar{\partial}_{\bar b}\Omega = - 2
\end{equation}
describes a Ricci-flat K\"ahler potential in a specific coordinate system.
It is only invariant  under antiholomorphic (or holomorphic)
coordinate transformations generated by divergence-free vector fields
with components
$v^{\bar a} = v^{\bar a} (\bar{Z})$. $\Omega$ transforms as a scalar,
\begin{equation}\label{gs}
\delta \Omega =  v^{\bar a} \bar{\partial}_{\bar a}
\Omega \;\;\;\text{with}\;\;\;
\bar{\partial}_{\bar a} v^{\bar a} =  0.
\end{equation}
However, the Plebanski equation possesses more symmetries. Consider
a vector field $\rho^{\bar a}(Z, \bar Z)\bar\partial_{\bar a}$ that
still has vanishing divergence but can depend on both $Z$ and $\bar Z$.
Then (\ref{pl2}) is invariant under a transformation $\delta\Omega$ satisfying
\begin{equation}\label{delta}
\partial_a \delta \Omega =  \rho^{\bar c} \bar{\partial}_{\bar c}
\partial_a \Omega.
\end{equation}
To check this statement we compute
\begin{eqnarray}
&&\epsilon^{ab} \epsilon^{\bar{a} \bar{b}} \partial_a \bar{\partial}_{\bar a}
\delta \Omega\, \partial_b \bar{\partial}_{\bar b}\Omega =
\epsilon^{ab} \epsilon^{\bar{a} \bar{b}} \bar{\partial}_{\bar a}
\big (\rho^{\bar c} \bar{\partial}_{\bar c}
\partial_a \Omega \big )
\, \partial_b \bar{\partial}_{\bar b}\Omega \nonumber \\
&& =  \epsilon^{ab} \epsilon^{\bar{a} \bar{b}} \bar{\partial}_{\bar a}
\rho^{\bar c} \bar{\partial}_{\bar c}
\partial_a \Omega \, \partial_b \bar{\partial}_{\bar b}\Omega +
\frac{1}{2} \rho^{\bar c} \bar{\partial}_{\bar c}\big (
\epsilon^{ab} \epsilon^{\bar{a} \bar{b}} \partial_a \bar{\partial}_{\bar a}
\Omega\, \partial_b \bar{\partial}_{\bar b}\Omega \big ) = 0.
\end{eqnarray}
The first term in the second line vanishes since
$\epsilon^{\bar{a} \bar{b}} \bar{\partial}_{\bar a} \rho^{\bar c}$ is
symmetric in $\bar b \leftrightarrow \bar c$
and the second term vanishes due to  the equation of motion.
For antiholomorphic components $\rho^{\bar a}(\bar{Z})$ the transformation
(\ref{delta}) reduces to (\ref{gs}) and corresponds to a diffeomorphism.
If however, the functions
$\rho^{\bar a}$  also depend on the holomorphic coordinates, i.e.
$\partial_b \rho^{\bar a} \neq 0$, the transformation (\ref{delta})
should not be thought of as due to a coordinate change of the underlying
complex manifold. It has no direct connection to diffeomorphisms.

Of course, this is not the full story. (\ref{delta}) is a differential
equation for $\delta \Omega$ which has a solution only if the right hand side
obeys the consistency condition
\begin{equation}\label{cc}
0 = \epsilon^{ab} \partial_a \partial_b \delta \Omega
=  \epsilon^{ab} \partial_a \rho^{\bar c} \bar{\partial}_{\bar c}
\partial_b \Omega.
\end{equation}
An infinite set of solutions $\rho^{\bar c}_n$, $n= 0, 1, 2, \ldots$ can
be found iteratively as follows. Let us assume there is  a solution
$\rho^{\bar c}_n$. Then one may verify that $\rho^{\bar c}_{n+1}$,
defined as
\begin{equation}\label{rn}
\partial_a \rho^{\bar c}_{n+1} = \bar{\partial}_{\bar a} \big (
\rho_n^{\bar b} \epsilon^{\bar{a} \bar{c}}\bar{\partial}_{\bar b}
\partial_a \Omega  \big ),
\end{equation}
is a new one. To complete the iterative description we
have to provide a vector field
$\rho_0^{\bar a}\bar\partial_{\bar a}$, which we take to be purely
antiholomorphic, i.e.  $\partial_a \rho_0^{\bar a}= 0$. This
evidently satisfies (\ref{cc}). The transformations generated by the
$\rho_n^{\bar a}$ via (\ref{delta}) are denoted by $\delta_n$ in the following.
Then only $\delta_0$ is a gauge symmetry as in (\ref{gs}), while the other
transformations are of the type (\ref{delta}).
It is important to note that the determination of the $\rho_n^{\bar a}$
according to (\ref{rn}) involves integrating the right hand side.
Thus $\delta_n$
is in general  a non-local symmetry \footnote{The iterative
construction  just described is in fact well known from the theory
of integrable models in two dimensions.  To see this rewrite
the Plebanski equation as $[W_0, W_1 ] = 0$ with $W_a$ from (\ref{va}).
Moreover, $W_a$ satisfies $\epsilon^{ab}\partial_a W_b = 0$. These
two equations are  the equations of a two dimensional model
derived from a  Wess-Zumino  action \cite{Pa}. Such models
are known to possess an infinite set of non-local symmetries that
are constructed as above \cite{S}.}.

As already mentioned in section two, in order to compare the results of this
section to the $N{=}2$ string we need to reformulate the symmetry structure
in terms of the field $\phi$, defined as
\begin{equation}
\Omega (Z, \bar Z) = \eta_{a\bar a} Z^a \bar{Z}^{\bar a} + \phi (Z, \bar Z).
\end{equation}
$\phi$ describes deviations of the K\"ahler potential from flat space.
For convenience we recall its equation of motion
\begin{equation}\label{pl3}
\eta^{a\bar a} \partial_a \bar{\partial}_{\bar a} \phi = \frac{1}{2}
\epsilon^{ab} \epsilon^{\bar{a} \bar{b}} \partial_a \bar{\partial}_{\bar a}
\phi \partial_b \bar{\partial}_{\bar b}\phi.
\end{equation}
The transformations $\delta_n $ read in terms of $\phi$:
\begin{equation}\label{dnp}
\partial_a \delta_n \phi =  \rho_n^{\bar c} \bar{\partial}_{\bar c}
\partial_a \Omega = \eta_{a\bar c} \rho_n^{\bar c} +
\rho_n^{\bar c} \bar{\partial}_{\bar c} \partial_a \phi,
\end{equation}
and one may check that they leave (\ref{pl3}) invariant. Moreover,
(\ref{rn}) becomes
\begin{equation}\label{rnp}
\partial_a \rho^{\bar c}_{n+1} = \epsilon^{\bar a}_a \bar{\partial}_{\bar a}
\rho_n^{\bar c} + \bar{\partial}_{\bar a} \big (
\rho_n^{\bar b} \epsilon^{\bar{a} \bar{c}}\bar{\partial}_{\bar b}
\partial_a \phi  \big )
\end{equation}
with $\epsilon^{\bar a}_a = \epsilon^{\bar a \bar b} \eta_{a \bar b}$.
It is instructive to work out the first few transformations explicitly:
\begin{eqnarray}
\delta_0 \phi &=& \eta_{a\bar c} Z^a \rho_0^{\bar c} +
\rho_0^{\bar c} \bar{\partial}_{\bar c} \phi, \nonumber \\
\rho_1^{\bar c} &=& \epsilon^{\bar a}_a Z^a \bar{\partial}_{\bar a}
\rho_0^{\bar c} + \epsilon^{\bar{a} \bar{c}} \bar{\partial}_{\bar a}
\big ( \rho_0^{\bar b} \bar{\partial}_{\bar b} \phi \big ),\nonumber \\
\partial_a \delta_1 \phi &=& \eta_{a\bar c} \epsilon^{\bar a}_b Z^b
\bar{\partial}_{\bar a} \rho_0^{\bar c} + \epsilon^{\bar a}_a
\bar{\partial}_{\bar a} \big ( \rho_0^{\bar c} \bar{\partial}_{\bar c} 
\phi \big )  + \epsilon^{\bar a}_b Z^b \bar{\partial}_{\bar a} \rho_0^{\bar c}
\bar{\partial}_{\bar c} \partial_a \phi \nonumber \\
&+& \epsilon^{\bar{a} \bar{c}}
\bar{\partial}_{\bar a}
\big ( \rho_0^{\bar b} \bar{\partial}_{\bar b} \phi \big )
\bar{\partial}_{\bar c} \partial_a \phi .
\end{eqnarray}
In general these   are  inhomogeneous  transformations, containing
$\phi$-independent terms. They are spontaneously broken by our
choice of ground state $\phi = 0$ which corresponds to a flat
background.
We cannot expect to see this kind of symmetry in first-quantised
string theory  since
there all symmetries are  connected to BRST invariant operators.
The latter can only generate unbroken symmetries, that do not contain
field-independent terms,  as is clear from the  discussion at the
end of subsection $4.1$. Yet if we take $\rho_0^{\bar c}$ to be constant,
i.e. consider global translations in the $\bar Z$ coordinates, it turns out
that the truncated transformation
\begin{equation}
\tilde{\delta}_0 \phi = \rho_0^{\bar c} \bar{\partial}_{\bar c} \phi
\end{equation}
still is a symmetry of the Plebanski equation since $\delta_0$ and
$\tilde{\delta}_0$ differ by a K\"ahler transformation.
Moreover, in $\rho_1^{\bar c}$ and $\delta_1$ the $\phi$-independent terms
disappear:
\begin{eqnarray}
\rho_1^{\bar c} &=& \epsilon^{\bar{a} \bar{c}}
\rho_0^{\bar b}\, \bar{\partial}_{\bar a} \bar{\partial}_{\bar b} \phi,
\nonumber \\
\partial_a \delta_1 \phi &=& \epsilon^{\bar a}_a
\rho_0^{\bar c} \bar{\partial}_{\bar c}
\bar{\partial}_{\bar a}\phi +
\epsilon^{\bar{a} \bar{c}} \rho_0^{\bar b}
\bar{\partial}_{\bar a} \bar{\partial}_{\bar b} \phi \,
\bar{\partial}_{\bar c} \partial_a \phi.
\end{eqnarray}
$\delta_1$  contains a linear and a  non-linear term. But the best we can do
is to compare the string transformations (\ref{trafo}) to the linearised
symmetries of the field $\phi$. We therefore focus in the following
on the linear part of the non-local transformations $\delta_n$.
Let us first note  that the iterative construction (\ref{rnp})
does not create
$\phi$-independent  terms in any  $\rho_n^{\bar c}$ derived from a
constant $\rho_0^{\bar c}$. This leads to the very important conclusion
that the corresponding transformations $\delta_n$ (one might call them
affinisations of translations) are unbroken symmetries.
Moreover, working to lowest order in $\phi$ we may drop the
second term on the right hand side of (\ref{dnp}). Differentiating
this equation $n-1$ times leads to
\begin{equation}
\partial_{a_1} \ldots \partial_{a_n} \delta_n \phi
= \partial_{a_1} \ldots \partial_{a_{n-1}} ( \eta_{a_n \bar{c}}
\rho_n^{\bar c})\;+\; {\cal O} ( \phi^2 ).
\end{equation}
Repeated application of (\ref{rnp}) and again dropping non-linear terms gives
\begin{equation}\label{linsym}
\partial_{a_1} \ldots \partial_{a_n} \delta_n \phi =
\epsilon^{\bar a_1}_{a_1} \ldots \epsilon^{\bar a_n}_{a_n}
\bar{\partial}_{\bar a_1} \ldots \bar{\partial}_{\bar a_n}
(\rho_0^{\bar c} \bar{\partial}_{\bar c} \phi )\; + \; {\cal O} ( \phi^2 ).
\end{equation}
After a Fourier transformation
\begin{equation}
\phi(Z,\bar{Z}) = {\int{d^4k}}\, e^{\frac{i}{2} (k{\cdot}\bar{Z} +
\bar{k}{\cdot}Z)}  \tilde{\phi} (k)
\end{equation}
we may impose the mass-shell condition $k \cdot \bar k = 0$.
Using  the relation
\begin{equation}
\eta_{a \bar a}
\bar{k}^{\bar a} = - h(k) \epsilon_{a b} k^b
\end{equation}
the  $n$-th transformation for $\tilde{\phi} (k)$ reads
\begin{equation}\label{trafo2}
\delta_n \tilde{\phi} (k) = \frac{i}{2} h(k)^{-n} \eta_{c \bar c}
\rho_0^{\bar c} k^c \tilde{\phi} (k) \;+\; {\cal O} (\tilde{\phi}^2).
\end{equation}
These are the non-local transformations derived from global translations
$\bar{Z}^{\bar a} \to \bar{Z}^{\bar a} + \rho_0^{\bar a}$.
The similarity to (\ref{trafo})
is evident since one may drop the constant prefactor $\frac{i}{2}
\eta_{c \bar c}\rho_0^{\bar c}$. We discuss in more detail what we think about
the relation of equations (\ref{trafo}) and (\ref{trafo2}) in section
six\footnote{
One may also compare this to equation $(4.1)$ from \cite{BV}.}.

Let us finally add that the unbroken linearised symmetries of equation
(\ref{pl3}) can be derived more quickly in the following fashion.
To find the linear part of a transformation
$\delta \phi$ one may simply drop the right hand side of
(\ref{pl3}) and study the equation
\begin{equation}\label{laplace}
\eta^{a\bar a} \partial_a \bar{\partial}_{\bar a} \delta \phi = 0.
\end{equation}
Starting from a rigid translation $\delta_0  \phi = \rho_0^{\bar c}
\bar{\partial}_{\bar c} \phi$ with constant $\rho_0^{\bar c}$ one can
iteratively define
\begin{equation}
\partial_a \delta_n \phi = \epsilon^{\bar a}_a \bar{\partial}_{\bar a}
\delta_{n-1} \phi
\end{equation}
which again leads to (\ref{linsym}).
Thus, finding the linearised symmetries amounts to the simple task
of solving (\ref{laplace}), e.g. by Fourier transformation.
Moreover, we see that different pictures correspond to different solutions
of (\ref{laplace}).  We have nevertheless prefered to
investigate the exact symmetries of the Plebanski equation in some more detail
in this section, hoping to better understand its connection to the
$N{=}2$ string.

\section{Ward identities}
Symmetries in string and field theories manifest themselves in relations
between correlation functions, known as Ward identities, which can be
extremely useful either for studying general properties of the theory or
for explicitly figuring out physically interesting quantities.
Therefore,  the Ward identities following from the symmetries
uncovered in  the previous section will now  be investigated.
Before turning to
the $N{=}2$ string, however, the first subsection will briefly review  the
general formalism  following \cite{WZ}, \cite{Kl} and \cite{V}.
\subsection{More generalities}
For simplicity we  restrict ourselves to lowest order
(genus zero and no $U(1)$ instantons) in string
perturbation theory in this paper, but higher orders can
be treated in a similar way. Moreover, in this subsection we
consider  bosonic string theory, in which perturbative
evaluation of correlation functions  involves only integration
over metric moduli. We defer to the next subsection the discussion of
how the more complicated
perturbative structure of  the  $N{=}2$ string affects the derivation of
Ward identities.

In the operator formalism a correlation function
\begin{equation}
\Theta = \big \langle V_1 \ldots V_{N+1} \big\rangle
\end{equation}
of $N+1$ BRST invariant operators $V_i$ with ghost numbers $g_i$ can be
regarded
as a differential form of degree $\sum_{i=1}^{N+1} g_i - 6$ on the moduli space
${\cal M}_{N+1}$ of the sphere with $N+1$ marked points (which is just the
configuration space of $N-2$ distinct points on the sphere).
This can be understood heuristically since $\sum_{i=1}^{N+1} g_i - 6$
is  the number of anti-ghost insertions and therefore of integrations needed
to obtain a number from $\Theta$.
Moreover, it is an important general result that this form is closed.
For scattering amplitudes one takes each of the $V_i$ to have ghost number
two. Then $\Theta$ is a top form and can be integrated over the full moduli
space to produce  a number -- the scattering amplitude.
To derive a Ward identity for the $N$-point function
one takes  $N$ physical vertex operators $V_i$
plus one ghost number one operator that is connected to the symmetries
of the theory. Then $\Theta$ is a form of codimension one and can only be
integrated over a submanifold of ${\cal M}_{N+1}$ of codimension one.
The natural candidate
is the boundary $\partial {\cal M}_{N+1}$. Because of $d\Theta = 0$ and
Stokes' Theorem this integral vanishes, so the Ward identity  reads
\begin{equation}
\int_{\partial {\cal M}_{N+1}} \Theta = \int_{{\cal M}_{N+1}} d\Theta = 0.
\end{equation}
The boundary of the moduli space of a sphere with $N+1$ marked points
corresponds to configurations where the sphere splits into two spheres
(one containing $N+1-p$ and  the other  $p$ of the points for
some $2\leq p\leq  N-1$) connected
by an infinitely  long tube. This tube can be represented by a complete set
of physical states propagating between the spheres. The twist angle of the
tube is one of the moduli leading to an insertion of
$b (z) - \tilde{b} (\bar{z})$. The complete set of states therefore takes
the form
\begin{equation}
\sum_i | \widehat{O}^i \rangle \langle O_i | ,
\end{equation}
where $i$ labels a basis of the absolute BRST cohomology and
\begin{equation}
\langle O_j |  O^i \rangle = \delta^i_j , \hs | \widehat{O}^i \rangle =
(b_0 - \tilde{b}_0) |  O^i \rangle .
\end{equation}
The Ward identity for a correlation function involving $N$ ghost number two
vertex operators $V_i := V(k_i, z_i , \bar{z}_i)$ and one ghost number one
operator $\Omega^{(0)}$ thus reads:
\begin{equation}\label{wi}
\sum_{i, \alpha}  \big\langle\! \big\langle V_{u_1} \hdots V_{u_p}
\Omega^{(0)}\widehat{O}^i
\big\rangle\! \big\rangle  \big\langle\!\big\langle  O_i V_{u_{p+1}}
\hdots V_{u_N} \big\rangle\!\big\rangle = 0.
\end{equation}
The sum over $\alpha$ runs over all possible ways to divide the $N$ physical
vertex operators into a subset $\{ V_{u_1} \hdots V_{u_p} \}$ on the sphere
of $ \Omega^{(0)}$ and the remainder $\{ V_{u_{p+1}} \hdots V_{u_N}\} $
located  on  the other sphere. Moreover, we have adopted the notation from
\cite{WZ} and indicate by a double bracket that the integration over
moduli space has already been performed.

\subsection{Ward identities of the $N{=}2$ string}
Before trying to apply the Ward identity (\ref{wi})  let us briefly comment
on the moduli of the $N{=}2$ string at tree level and zero $U(1)$  instanton
number. The integration over the fermionic moduli can be explictly
performed, resulting in the prescription that the picture numbers $\pi^+ =
\tilde{\pi}^+$ and $\pi^- = \tilde{\pi}^-$ of the vertex operators
both have to add up to $-2$ in order to get a non-vanishing correlator.
The $U(1)$ moduli enter via the homology  of the punctured
Riemann surface. At genus zero they simply correspond to the
non-trivial monodromies of the $U(1)$ charged fields around the punctures.
However,  spectral flow identifies the correlators for different monodromies
and we can restrict ourselves to the NS sector at each puncture.
The integration over the $U(1)$ moduli merely contributes to the normalisation.
Moreover, the $U(1)$ moduli space has no boundaries. It can  play
no role in this subsection as one sees from  the way the Ward identity
has been derived above. We will therefore not  write down
the  $U(1)$ ghosts explicitly.

Let us now consider the Ward identities involving the ghost number one
states (\ref{coh2}) and suppress  the picture numbers for notational
simplicity. For the case of the three-point function involving
three closed-string vertex operators $V_i (k_i)$ with lightlike
momenta $k_1 + k_2 + k_3 = 0$ the identity (\ref{wi}) becomes\footnote{For
the three-point function on the sphere there is no difference between single
or double brackets since the corresponding moduli space consists of a single
point.}
\begin{align}
&\sum_i \big \langle V_1 \Omega^{a(0)}_{m,n}\widehat{O}^i
\big \rangle \big \langle  O_i V_2 V_3 \big \rangle +
\sum_i  \big \langle V_2 \Omega^{a(0)}_{m,n}\widehat{O}^i
\big \rangle \big \langle  O_i V_3 V_1 \big \rangle \nonumber \\
&+ \sum_i  \big \langle V_3 \Omega^{a(0)}_{m,n}\widehat{O}^i
\big \rangle \big \langle  O_i V_1 V_2 \big \rangle = 0.
\end{align}
The picture and ghost numbers and the momenta of the operators $O_i$ and
$\widehat{O}^i$ are uniquely fixed. For example, the only operator that
contributes
in the first sum is $O_i = V_1$ with conjugate states $O^i= \partial c
\bar{\partial} \tilde{c} V_1$ and $\widehat{O}^i = (\partial c +
\bar{\partial} \tilde{c}) V_1$. The second factor in the first term
is simply the ordinary
three-point function $A_3 $, while the first factor
splits into holomorphic and antiholomorphic parts
\begin{align}\label{cor}
&\big \langle V_1 (k_1) \Omega^{a(0)}_{m,n}  (\partial c +
\bar{\partial} \tilde{c}) V_1(-k_1) \big \rangle \nonumber \\
& = \big \langle V_1(k_1) {\cal P}^a_m  V_1(-k_1) \big \rangle_L \big \langle
\tilde{V}_1(k_1) \tilde{{\cal O}}_n \bar{\partial} \tilde{c}\tilde{V}_1(- k_1)
\big \rangle_R.
\end{align}
The subscripts $L$ and $R$ indicate that the two correlators are
meant with respect to the chiral left- and right-moving theories
(the term $\partial c V_1$
does not contribute because of ghost number counting).
Note that the whole
discussion does not depend on the pictures chosen, so that there is no
restriction on the values of $m$ and $n$. The correlators in (\ref{cor}) can
be evaluated using the relations from section $3$ and then  yield
$k_1^a h(k_1)^{m+n}$.
The explicit Ward identity for the three-point function finally takes the form
\begin{equation}
A_3  \sum_{i=1}^3 k_i^a h(k_i)^m = 0
\end{equation}
for any $m \in\mathbb{Z}$.
An analogous relation holds with $k_i^a$ replaced by $\bar{k}_i^{\bar a}$.
The amplitude can  be non-zero only if the sum above  vanishes for all $m$.
The unique solution is
\begin{equation}\label{w3}
h(k_1) = h(k_2) = h(k_3).
\end{equation}
This exactly coincides with the result of Berkovits and Vafa in \cite{BV} to
which we refer for  further discussion.

Next, we discuss the Ward identities for general $N$-point functions involving
$N$ lightlike momenta $k_i$ with $k_1 + k_2 + \hdots  + k_N= 0$.
The $\alpha$ sum in (\ref{wi}) now runs over a large number of possible
degenerations of the Riemann surface.
In particular, in the splitting  the ghost number zero operator
$\Omega^{(0)}$ may be accompanied by  more than one vertex operator,
leading to a correlator
\begin{equation}
\big\langle\!\big\langle V_{u_1} V_{u_2} \hdots V_{u_p}
\Omega^{a(0)}_{m,n}\widehat{O}^i
\big\rangle\!\big \rangle \big\langle\!\big\langle   O_i V_{u_{p+1}}
\hdots  V_{u_N} \big\rangle\!\big\rangle.
\end{equation}
$O_i$ has momentum $k_{u_1} + k_{u_2}+ \hdots +k_{u_p}$ and is generally
not an on-shell vertex operator. The  standard way to deal with  this
situation is to invoke the canceled propagator argument: Evaluate the
above expression in a kinematical region where the intermediate states
have positive scaling dimension. The contributions from the boundary
of moduli space vanish in this case. By virtue of analytic continuation
the correlator must then  also vanish in other regions of momentum space.
However, if $\Omega^{(0)}$ splits off with only one vertex operator $V_i$
the canceled propagator argument does not apply since the intermediate
state in this situation has momentum $k_i$ and is always on-shell.
The Ward identity therefore receives contributions only from this
type of degeneration of the Riemann surface and reads
\begin{equation}\label{win}
\sum_{i=1}^N \langle V(k_i) \Omega^{a(0)}_{m,n} \widehat{V}(-k_{i})\big\rangle
A_N  = A_N  \sum_{i=1}^N k_i^a h(k_i)^{m+n} = 0
\end{equation}
which holds for any $m+n \in \mathbb{Z}$. It does not
look very exciting (after all, it stems from affinisations of
translations) but implies that $A_N (k)$ vanishes unless
\begin{equation}\label{wim}
\sum_{i=1}^N k_i^a h(k_i)^m = 0\hs \text{for any $m$} \in \mathbb{Z}.
\end{equation}
To study the solutions of this equation it is useful to recall
that $h$ is only a phase and can be rewritten as $h(k_i) = e^{i \gamma_i}$.
Dividing (\ref{wim}) by $h(k_1)^m$ and summing over $m$ leads to
\footnote{We are grateful to Helge Dennhardt for this suggestion.}
\begin{equation}\label{sum}
0 = \sum _{i=1}^N k_i^a \sum_{m \in \mathbb{Z}} e^{im (\gamma_i - \gamma_1)} =
\sum _{i=1}^N k_i^a \delta ( \gamma_i - \gamma_1).
\end{equation}
The first term, $i=1$, in the sum is non-vanishing.
Without loss of generality  we can neglect kinematical situations
where a true subset of the
momenta sums to zero, so the only possibility  for (\ref{sum}) to hold is
\begin{equation}
\gamma_1 = \gamma_2 \hdots =  \gamma_N.
\end{equation}
Then (\ref{sum}) is satisfied because of momentum conservation.
The Ward identity
(\ref{wim}) therefore leads to the final conclusion that the $N$-point
function vanishes unless
\begin{equation}
h(k_1) = h(k_2) = \hdots =h(k_N)
\end{equation}
which implies that all scalar products $k_i{\cdot}\bar{k}_j +k_j{\cdot}
\bar{k}_i$ are zero
and  exactly reproduces the general vanishing theorem for tree-level
amplitudes of \cite{BV}.

In the discussion above we neglected special kinematical situations
and invoked analytic continuation.
In a  realistic theory  this would certainly be justified.
Whether or not one should require on-shell correlation functions to be
analytic in a spacetime of signature $(2,2)$ depends on one's
interpretation. For example, Parkes argues in \cite{Parkes} that
$\delta$-function contributions to the $S$-matrix of the $N{=}2$ string are
important with respect to the role of self-dual gravity and $N{=}2$ string
theory in the theory of integrable models. Clearly, such subtleties
are not accessible by the above methods.

\section{Concluding remarks}

The  purpose of this paper is to provide evidence for our
belief that the non-trivial picture structure of the BRST cohomology
is not just an irrelevant detail of the BRST approach
but important for a deeper  understanding  of the theory
(see \cite{BZ} for similar remarks). The key points are:
\begin{itemize}
\item
The symmetry transformation  (\ref{trafo})  of the
first-quantised string theory coincides with the symmetry transformation
(\ref{trafo2}) of the field theory  at the linearised  level.
\item
The Ward identity (\ref{wim}) correctly implies the vanishing of all
correlation functions with more than three external legs, at least for
generic values of the momenta.
\end{itemize}
The first point deserves further discussion. It is, of course, not a simple
task to discover  the full symmetry group of a string model. Doing so
would roughly correspond to having found a useful non-perturbative definition
of the theory. In this paper we have worked in the standard first-quantised
formalism. To see how much insight into the symmetry structure one may gain
within such an approach, it is instructive to recall
the situation in closed bosonic string theory. The latter includes
gravity and should therefore be a theory invariant under general coordinate
transformations. The standard perturbative expansion, however, is around flat
space which breaks the symmetry down to the Poincar\'e group.
The zero momentum cohomology at ghost number one contains  only
translations (see (\ref{bosstr})). One can certainly write down
currents that generate Lorentz transformations,
\begin{equation}\label{lorgen}
J^{\mu\nu} = X^{\mu} \partial X^{\nu} - X^{\nu} \partial X^{\mu},
\end{equation}
but they do not show up in the cohomology since $X^{\mu}$ is not a legal
operator that takes part in the operator-state correspondence of the
conformal field theory.

Equivalently, the field $\phi$ that encodes the degrees of freedom  of the
$N{=}2$ string describes deviations from  the K\"ahler potential
of flat space. Hence  we find  translations in the $(0,0)$ picture
(see (\ref{mom})).
This is the most natural picture in the sense that it is the only one
where currents derived from cohomology classes do not contain contributions
from the ghost sector. They can thus be integrated over the world sheet
to generate deformations of the conformal field theory.

We have  seen in subsection $4.3$ that the  {\it non-local}
transformations derived from
translations also leave the flat metric invariant. We should therefore
find them somewhere in the zero momentum cohomology of the $N{=}2$ string.
The similarity between equations (\ref{trafo}) and (\ref{trafo2})
suggests that they show up via the picture dependence of the zero
momentum BRST cohomology, but it is needless to say that at the
moment this is just a tentative statement.
There are further unbroken symmetries in the field theory. One can check,
for example, that the non-local transformation $\delta_2$ derived
from a rotation, i.e. $\rho_0^{\bar c} = \bar{Z}^{\bar c}$,
is free of  field-independent parts. However, the target space coordinates
 appear  explicitly in the transformation law, and we should not
expect to find a counterpart in the cohomology -- just as
explained above in the context of bosonic string theory.

One further aspect should be mentioned. The transformations $\delta_n$
are gauge symmetries only for $n=0$. On the other hand, it is
a general theorem
that string theory does not admit  continuous rigid
symmetries in target space (see chapter $18$ of \cite{Pol}). The proof of this
statement is based on the fact that for any such symmetry there are
associated conserved currents  on the world-sheet from which one can
construct vertex operators for gauge bosons. One may therefore wonder
how the unbroken  non-local symmetries can be realised in $N{=}2$
string theory. The point is that the above theorem typically assumes
that the world-sheet currents are chirally conserved, which is not always
true. One counter example are the currents (\ref{lorgen}) of
Lorentz transformations (`string theory gauges only translations', \cite{BL}).
In the case of the $N{=2}$ string the currents  associated to the  symmetry
charges are given in (\ref{o1}). They are chirally conserved only
for $(\pi^+,\pi^-) = (0,0)$ since then the right moving part is just the
unit operator. The corresponding gauge boson is the Plebanski field $\phi$.
For other values of $(\pi^+,\pi^-)$ with  $m=n=0$ in (\ref{o1})
the currents are not chirally conserved but nevertheless
give rise to gauge bosons  which are
picture changed versions of the $(0,0)$ vertex operator.
For non-zero values of $m$ and $n$, however, the currents do not allow one
to construct a gauge boson vertex operator, which explains
why they correspond only to global symmetries in target space.

It is also interesting to note the similarity between the $N{=}2$ string
and  $2D $ strings:
Both theories have the same spectrum (one scalar) and
possess a ring of ghost number zero states. This leads in both
cases to an infinite dimensional symmetry which completely fixes the dynamics.
The ground ring elements  of the $2D$ theory are labelled by 
quantised values of the momenta. On physical vertex operators
(tachyons) they act by muliplication of a momentum dependent function
and by a shift of the momenta. This is completely analogous to the relation
$$
{\cal O}_{\pi^+,\pi^-, n} \cdot V_{\widehat{\pi}^+,\widehat\pi^-} (k) = h(k)^n
V_{\pi^+ + \widehat{\pi}^+,  {\pi}^- + \widehat{\pi}^-} (k),
$$
which  suggests that the fields $\varphi^{\pm}$ appearing in the
bosonisation of the spinor ghosts  play a  role similar to the
Liouville field in two dimensions. Both  are scalar fields
coupled to a background charge, and their momenta are quantized.
For the $N{=}2$ string this suggests the appearance of an additional
complex dimension.  Such extra dimensions in connection with
the $\beta$, $\gamma$ ghost system have also been proposed in \cite{B1}.

Finally, it would be interesting to find further examples in which
the picture structure yields non-trivial information about a theory.
In \cite{BZ} the picture dependence of the relative  zero-momentum
cohomology of the Ramond sector of the $N{=}1$ string in flat space has been
discussed.  Using the picture independence of the absolute cohomology it is, 
however,  not hard to show
that no new ghost number zero cohomology classes appear in higher pictures
(we have checked this for the $1/2$ and $3/2$  picture, and
it seems unlikely that new states appear at still higher pictures).
To find such a  phenomenon in  a $10D$ superstring theory  it is certainly
necessary to consider non-trivial backgrounds.
For example, in \cite{GKS} an investigation of string propagation
on a manifold that includes $AdS_3$ has been initiated.
Because of the recently proposed $CFT/AdS$ correspondence an infinite
symmetry is expected in this type of theory, which might possibly show up in
some cases through a picture dependence of the BRST cohomology.

\vspace{1.0cm}
\noindent
{\large\bf Acknowledgements}\\
\noindent
K.J. would like to thank F. Brandt, H. Dennhardt  and B. Niemeyer
for useful discussions.
A.D.P. thanks the Institut f\"ur Theoretische Physik der
Universit\"at Hannover for its hospitality. The work of A.D.P.
was partially supported by the Heisenberg-Landau Program.

\vfill\eject

\appendix
\section{Zero-momentum states}
For convenience we summarise in this subsection the chiral
zero-momentum cohomology
classes of ghost number zero and one in various pictures:

\vspace{0.5cm}
\begin{center}
\begin{tabular}{|c|c|c|}
\multicolumn{3}{c}{Ghost number zero}\\ \multicolumn{3}{c}{}\\
\hline
picture number & cohomology classes & dimension of \\
$(\pi^+,\pi^-)$  & & cohomology  \\
\hline
$(-2-p,p)$ & ---  & 0 \\
$(-1-p,p)$ & --- & 0  \\
$(-p,p)$ & $A^p$ & 1 \\
$(1-p,p)$ & $X^+ \cdot A^p , \hs X^- \cdot A^{p-1}$ & 2 \\
$\pi^+ + \pi^- > 1$ & ${\cal O}_{\pi^+,\pi^-,n} := 
(X^+)^{\pi^++n}\cdot (X^-)^{\pi^--n}\cdot A^n,$ & $\geq \pi^+ + \pi^- +1$\\
& for $n=- \pi^+ \hdots \pi^-$ & \\
& plus possibly more states.& \\
\hline
\end{tabular}
\end{center}

\noindent
Here $p\in \mathbb{Z}$ is an arbitrary integer, the normal
ordered product is defined in (\ref{mult}), and the explicit
expressions for the operators $A$, $A^{-1}$
and $X^{\pm}$ are
\begin{align}
A &=  (1 - c b') J^{--} e^{\varphi^+} e^{-\varphi^-},\nonumber \\
A^{-1} &= (1 + c b') J^{++} e^{-\varphi^+} e^{\varphi^-},\\
X^{\pm} &= - c \partial \xi^{\pm} + \big ( G^{\pm} - 4
\gamma^{\pm} b \pm 4 \partial \gamma^{\pm} b' \pm 2 \gamma^{\pm} \partial
b'\big )  e^{\varphi^{\mp}}.\nonumber
\end{align}

\vspace{0.5cm}
\begin{center}
\begin{tabular}{|c|c|c|}
\multicolumn{3}{c}{Ghost number one}\\ \multicolumn{3}{c}{}\\
\hline
picture number & cohomology classes & dimension of \\
$(\pi^+,\pi^-)$  & & cohomology \\
\hline
$(-2-p,p)$ & $A^{p+1} \cdot c e^{-\varphi^+} e^{-\varphi^-}$ & 1 \\
$(-1-p,p)$ & $A^p \cdot c\psi^{+a}  e^{-\varphi^-}$  & 2 \\
$(-p,p)$ & $A^p \cdot {\cal P}^a, \; A^p \cdot \bar{\cal P}^{\bar a} $ & 4 \\
$(1-p,p)$ & $X^+ \cdot A^p \cdot {\cal P}^a ,
\hs  X^- \cdot A^{p-1}\cdot {\cal P}^a$ & $\geq 8$  \\
& $X^+ \cdot A^p \cdot \bar{\cal P}^{\bar a} ,
\hs X^- \cdot A^{p-1}\cdot \bar{\cal P}^{\bar a}$ & \\
& plus possibly more states.& \\
$\pi^+ + \pi^- > 1$ & ${\cal O}_{\pi^+,\pi^-,n} \cdot {\cal P}^a,\hs
{\cal O}_{\pi^+,\pi^-,n} \cdot \bar{\cal P}^{\bar a}$ &
$\geq 4 ( \pi^+ + \pi^- +1)$\\
& plus possibly more states. &\\
\hline
\end{tabular}
\end{center}

\noindent
The momentum operators are
\begin{align}
-i{\cal P}^a \equiv  -i{\cal P}^a_{0,0} &= 
c \partial Z^{a} - 2 \gamma^- \psi^{+a}, \nonumber \\
-i\bar{\cal{P}}^{\bar a} \equiv  -i \bar{\cal{P}}^{\bar a}_{0,0} &=
c \partial \bar{Z}^{\bar a} - 2 \gamma^+ \psi^{-\bar a}.
\end{align}
Due to Poincar\'e duality one obtains  similar tables for
$\pi^+ + \pi^- \leq -2$.

\section{The ghost system}
Only the currents in (\ref{n2currents}) arise as constraints from gauge-fixing
the $N{=}2$ supergravity theory on the world-sheet. They have to be used to
construct the chiral  BRST operator. We furthermore  need the standard
$b$,$c$ ghosts of weight $(2,-1)$,
spinor ghosts $\beta^{\pm}$, $\gamma^{\mp}$ of weight $(3/2,-1/2)$ and
$U(1)$ ghosts $b'$, $c'$ of weight $(1,0)$. Their mode expansions
and commutation relations are standard (the spinor ghosts are
half-integer moded since we work in the NS sector).
The ground state $|\; \pi^+,\pi^- \;\rangle$ with picture number
$(\pi^+,\pi^-)$
is defined by dividing the spinor ghost modes into annihilators and creators:
\begin{align}
\beta_r^{\pm} |\;  \pi^+,\pi^-\;\rangle &= 0 \;\; {\rm when}\; \;
r \geq - \pi^{\pm}-\frac{1}{2}, \nonumber \\
\gamma_r^{\mp} |\;  \pi^+,\pi^-\;\rangle &= 0 \;\; {\rm when}\; \;
r \geq \pi^{\pm} + \frac{3}{2}.
\end{align}
It is useful to bosonise the spinor ghosts:
\begin{equation}
\gamma^{\pm} \rightarrow \eta^{\pm} e^{\varphi^{\pm}}, \hs
\beta^{\pm} \rightarrow e^{-\varphi^{\mp}} \partial \xi^{\pm}.
\end{equation}
$\eta^{\pm}$ and $\xi^{\pm}$ are fermionic fields with weight $1$ and $0$
respectively  and
OPE $\eta^{\pm}(z) \xi^{\mp} (w) \sim (z-w)^{-1}$. $\varphi^{\pm}$ are bosonic
scalars that couple to a background charge. Their OPE is $\varphi^{\pm}(z)
\varphi^{\pm}(w) \sim - \ln(z-w)$.
With these variables the state $|\; \pi^+,\pi^- \;\rangle$ can be created
from the $sl(2)$ invariant ground state $|0\rangle$ as
\begin{equation}
|\; \pi^+,\pi^- \;\rangle = e^{\pi^+ \varphi^- +\pi^- \varphi^+}(0) |0\rangle.
\end{equation}
As in \cite{B} we define the ghost number current in a slightly unusual way as
\begin{equation}
j_{gh} = - bc - b' c' + \eta^+ \xi^- + \eta^- \xi^+.
\end{equation}
This assigns the correct ghost number to all ghost fields, but commutes
with the operators $X^{\pm}$ and $A$ defined in section $3$. Moreover, all
states  $|\; \pi^+,\pi^-  \;\rangle$ have zero ghost number.

The chiral  BRST operator $Q=Q^{mat} + Q^{gh}$ splits into two pieces:
\begin{align}
Q^{mat} &= \sum_n\big (  c_{-n} L_n + c'_{-n} J_n \big )
+ \sum_r\big(\gamma^+_{-r} G^-_r + \gamma^-_{-r} G^+_r \big ), \nonumber \\
Q^{gh} &=  -\frac{1}{2} \sum (m-n) :c_{-m} c_{-n} b_{m+n}: -
\sum m :c'_{-m} c_{-n} b'_{m+n}: \nonumber \\
&+ \frac{1}{2} \sum (n-2s)c_{-n} 
:(\gamma^-_{-s}\beta^+_{n+s}  +  \gamma^+_{-s} \beta^-_{n+s} ): \nonumber \\
&- \sum c'_{-n} :(\gamma^+_{-s}\beta^-_{n+s} - \gamma^-_{-s} \beta^+_{n+s}):
\nonumber \\
&- 4 \sum \gamma^-_{-s} \gamma_{-r}^+ b_{r+s} - 
2\sum (s-r) \gamma^-_{-s} \gamma_{-r}^+ b'_{r+s}.
\end{align}
Normal ordering is defined with respect to the $sl(2)$ invariant ground state,
i.e. the spinor ghosts are normal ordered with respect to the $(0,0)$ picture.
The ghost parts of the super Virasoro generators can be obtained by
anticommuting $Q$ with the anti-ghost modes. The explicit expressions for
$L^{gh}_0$ and $J^{gh}_0$ are
\begin{align}
L^{gh}_0 &= \sum_m m :c_{-m} b_{m}: + \sum_m m  :c'_{-m} b'_{m}:
- \sum_s s :(\gamma^-_{-s}\beta^+_{s} +  \gamma^+_{-s}\beta^-_{s}):,
\nonumber\\
J^{gh}_0 &= - \sum_s :(\gamma^+_{-s}\beta^-_{s}-  \gamma^-_{-s}\beta^+_{s}):.
\end{align}
On the states $|\; \pi^+,\pi^- \;\rangle$ these operators act as
\begin{align}
L_0^{gh} |\; \pi^+,\pi^-  \;\rangle &= 
\Big [  \pi^+ \Big (\frac{\pi^+}{2} + 1 \Big )+ \pi^- \Big (\frac{\pi^-}{2} + 
1 \Big ) \Big ] |\; \pi^+,\pi^- \;\rangle, \nonumber \\
J_0^{gh} |\; \pi^+,\pi^-  \;\rangle &= \big ( \pi^- - \pi^+ \big )
|\; \pi^+,\pi^-  \;\rangle.
\end{align}
These equations are useful to explicitly construct states in the  relative
Fock space (\ref{rfs}).

\end{document}